\documentclass[aps,pra,twocolumn,superscriptaddress,showpacs]{revtex4}

\usepackage{latexsym}
\usepackage{epsfig} 
\usepackage[utf8]{inputenc}
\usepackage[english]{babel}
\usepackage{amsmath}
\usepackage{amsfonts}
\usepackage{amssymb}
\usepackage{graphicx}
\usepackage{theorem}

\usepackage{bbm}

\newcommand{\emes}{\mathcal{E}_{\mathcal{M}}}
\newcommand{\lmes}{\mathcal{L}_{\mathcal{M}}}
\newcommand{\eof}{\mathcal{E}_{\!O\!F}}

\newcommand{\rw}{\rho_{W}}
\newcommand{\ri}{\rho_{I}}
\newcommand{\trace}[1]{\mbox{Tr} \left[ #1 \right]}
\newcommand{\tw}{\mathcal{P}_\Lambda}
\newcommand{\foo}{b}
\newcommand{\fopt}{\foo_{\text{opt}}}
\newcommand{\fwer}{\foo_{\text{Wer}}}
\newcommand{\fiso}{\foo_{\text{Iso}}}
\newcommand{\id}{\mathbbm{1}}
\newcommand{\fid}{\mathbb{F}}
\newcommand{\gid}{\mathbb{G}}

\newcommand{\qid}{\mathbb{Q}}

\providecommand{\be}{\begin{eqnarray}}
\providecommand{\enn}{\end{eqnarray}}

\begin{document}

\title{Quantitative entanglement witnesses of Isotropic- and Werner-class via local measurements} 

\author{P. Silvi}
\affiliation{International School for Advanced Studies (SISSA), Via Bonomea 265, 34136 Trieste , Italy}
\author{F. Taddei}
\affiliation{NEST, Scuola Normale Superiore and Istituto Nanoscienze-CNR, Piazza dei Cavalieri 7, I-56126Pisa, Italy}
\author{R. Fazio}
\affiliation{NEST, Scuola Normale Superiore and Istituto Nanoscienze-CNR, Piazza dei Cavalieri 7, I-56126Pisa, Italy}
\author{V. Giovannetti}
\affiliation{NEST, Scuola Normale Superiore and Istituto Nanoscienze-CNR, Piazza dei Cavalieri 7, I-56126Pisa, Italy}
\date{\today}

\begin{abstract}

Quantitative entanglement witnesses allow one to bound the entanglement present in a system by acquiring a single expectation value. 
In this paper we analyze a special class of such observables which are associated with (generalized) Werner and Isotropic states. For them the optimal bounding functions can be easily derived by exploiting known results on twirling transformations. By focusing on an explicit local decomposition for these observables we then show how simple classical post-processing of the measured data can tighten the entanglement bounds. Quantum optics implementations based on hyper-entanglement generation schemes are analyzed.

\end{abstract}

\maketitle

\section{Introduction} \label{sec:intro}

Detecting and characterizing  entanglement in  quantum systems is
a major issue of quantum information theory,  both from the theoretical  and from the experimental point of view \cite{PREP,bruss}.
In this context
Entanglement Witnesses (EWs)   turn out to be useful tools that allow
one to 
address this  problem   with a minimal experimental effort, even in those cases where a complete  tomographic reconstruction of the  state  is not experimentally accessible \cite{EWIT}. 
Indeed, this technique merely requires one to measure a single (non local)  observable whose expectation value yields  a certifiable
evidence of the presence of entanglement in the system. 
 Unfortunately it  cannot provide a definitive answer for all system configurations
(indeed  in most of the cases, the result of the measurement will be inconclusive). Furthermore, 
even when successful, the result of an EW  measure will not
be useful in general to quantify the amount of entanglement present in  the system. Still it was recently
 pointed out \cite{GHUNE1,GHUNE2,OST,AUDE} that the outcome of such measurements  can be used to provide at least 
non trivial  (lower) bounds for the latter. 
These observations led to the notion of 
Quantitative Entanglement Witness (QEW) \cite{EIS}, which identify those operators
whose expectation values can be translated into a quantification of the entanglement present in the measured
system. 

From a purely technical  point of view, it is worth stressing that even though 
QEWs were originally discussed in the context of standard EWs,  they can be  thought as independent from the latter. 
Indeed an EW is characterized by well established properties, e.g.  it must assume positive values on 
all separable states, and it must have at least one negative eigenvalue \cite{EWIT}. On the contrary 
the definition of QEW can be applied to any operator $\qid$ for which it is
possible to associate a non trivial {\em bounding function} (see next section) for a given entanglement measure.
Notable examples of such  QEWs were already known in literature even before the formal introduction of this notion, e.g.
see \cite{Albe}.
In particular  the idea of bounding  the entanglement of a state via the acquisition of a single 
expectation value was addressed in a series of papers \cite{burkardloss} that applied previous
ideas on twirling transformations \cite{BENNETT} and Werner states \cite{VOLL,WERNER};
the same argument has been extended lately to Isotropic states \cite{Albe,Brand,HORO,TERHAL,VIDAL,RUNGTA,VOLL,LEE}.
The geometrical properties of Werner and Isotropic states can be used to build explicitely QEW
capable of detecting up to a high degree of bipartite entanglement.
In this paper we review the derivation of such QEW \cite{EIS,GHUNE1,GHUNE2,OST}
adopting a formal yet simple scheme.
Subsequently using an explicit decomposition  of such operators in terms of correlated local observables \cite{MACCH1}  we perform a data-processing optimization
which will enable further tightening of the entanglement bounds.
Expressing EW or QEW in terms of local operation and classical communication (LOCC) measurement schemes
is  yet another important aspect of entanglement detection 
\cite{MACCH1,MACCH,SAN,TERHAL2,PITTENG, FILIP, PREP, THESIS, Albe,Brand}. It   deals with 
the problem of acquiring expectations values of a non-local observable (i.e. an observable that does not admit a separable basis of eigenvectors),
when only local resources are  accessible experimentally. Even though the  decomposition we assume  is not the most efficient one in terms of setup preparations \cite{PREP},
it has the main advantage of being easily implementable in  some specific experimental configurations which exploit multi-rail quantum encoding strategies
\cite{HYP1,HYP,HYP2,GAO,multi,beenakker,BEEN11}.
We will explicitly focus on some optical scheme based on {\em hyper-entanglement} setups \cite{HYP,GAO,HYP1} where bipartite entangled qudits states are generated by exploiting
spatial and polarization degrees of freedom. 

The paper is organized as follows.
After reviewing some basic definitions around quantitative entanglement witness, in section \ref{sec:QEW} we focus on 
QEWs of Werner class and of Isotropic class for which we present the optimal bounding  functions.  In Sec. \ref{sec:deco} we introduce the local decompositions for such QEWs which use
$2d$ independent set-up preparations ($d$ being the local dimension of a system site). We then show how  the same  measurement outcomes 
can  be post-processed  to strengthen the entanglement bounds. Sec. \ref{sec:AS} finally discusses how to implement the proposed measurement scheme in the context of optical multi-rail encodings.
\section{Quantitative entanglement witnesses} \label{sec:QEW}

Let $\qid$ be an observable acting on a bipartite system $AB$, and let
$\emes$ be a given convex and continuous measure of bipartite entanglement, e.g. the entanglement of formation \cite{BENNETT,PREP,bruss}.
We say that $\qid$ is a QEW for the measure $\emes$  if we can define a (nontrivial)
non-negative 
function $\foo$ for which it holds
\begin{equation} \label{eq:qew}
 \emes \left( \rho\right) \geq
 \foo \left( \trace{\qid \,\rho} \right) \;,
\end{equation}
for all $\rho$ density matrices of $AB$.
Analogously to a standard entanglement witness \cite{PREP,EWIT,bruss}, a QEW provides an operational method for detecting entanglement. Furthermore it also
yields a quantitative estimation of it. 
As pointed out in \cite{EIS, GHUNE1,GHUNE2}, 
a fundamental problem 
 is to find the \emph{optimal} bounding function 
of a given QEW $\qid$, i.e. the highest positive $\foo$ for which \eqref{eq:qew} holds.
The latter can be formally defined as 
\begin{equation} \label{eq:optimal}
 \fopt (x) = \inf_{\rho} \left\{ \emes(\rho) \left|\; \trace{\qid \,\rho} = x \vphantom{\sum} \right.\right\}\;,
\end{equation}
where  the minimization is performed over the whole set of density matrices. This is a 
convex (generally not monotonic) function which typically is rather difficult to obtain. 
A partial characterization of $\fopt$ was proposed in \cite{GHUNE2} where it was pointed out that,
due to the convexity of entanglement measures, we can write
\begin{equation} 
 \fopt (x) = \sup_{\lambda} \left\{ \lambda x - \lmes(\lambda \qid)\right\}\;, \label{alt}
\end{equation}
where the supremum is taken over all real values of  the parameter  $\lambda$ and where $\lmes$ is the
{\em Legendre transform} of the measure functional $\emes$, with respect to the trace scalar product, i.e.
\begin{equation}
 \lmes (A) \equiv \sup_{\rho} \left\{ \trace{A \,\rho} - \emes(\rho) \right\}\;,
\end{equation}
(here the supremum is  taken over the whole set of density matrices). 

A useful property of the set of QEW is that it is invariant under Local Operation and Classical Communication (LOCC).
In particular any (non trivial) unitary LOCC   transformation $\Gamma$ takes a given QEW
into a new QEW  while  preserving its bounding functions including the optimal one, which is trivially again optimal \cite{NOTA11}.
This is a simple consequence of  the cyclicity of the trace which gives
\begin{equation} \label{eq:questa} 
 \emes(\rho) = \emes \left( 	\Gamma \,(\rho)\, \right) 
 \geq \foo \left( \trace{\Gamma^{\ast}( \qid )\, \,\rho} \right),
\end{equation}
for all $\foo$ of $\qid$ (in this expression $\Gamma^{\ast}$ stands for the adjoint super-operator of $\Gamma$). 
Notice that exploiting the properties of the transformations $\Gamma$, we can derive a stronger bound on $\emes(\rho)$, namely \cite{AUDE}
\begin{equation} \label{eq:questa1} 
 \emes(\rho) 
 \geq \max_{\Gamma} \; \foo_{\text{opt}}  \left( \trace{\Gamma^{\ast}( \qid )\, \,\rho} \right)
 = \foo_{\text{opt}} (q_*) \;,
\end{equation}
where the maximization is now performed over the set of all LOCC unitaries, and where $q_*$ indicates  the value of $\trace{\Gamma^{\ast}( \qid )\, \,\rho}$ that yields the highest value of 
$\fopt$. Unfortunately  $q_*$ is  a rather complicated (nonlinear) function  of the state $\rho$ which is arguably 
hard to compute. Still, we will see in the following that, for the cases we consider here, an operational
characterization of it can be obtained by focusing on a subclass of $\Gamma$.

\subsection{Werner and Isotropic QEWs}

Finding  the optimal bounding functions for  QEWs is a hard
problem \cite{EIS,GHUNE1,GHUNE2}. However  there are certain highly symmetric choices of $\qid$ 
for  which $b_{\text{opt}}$ can be obtained without going through the optimizations of (\ref{alt}). 
Prototypical examples are given by the operators
\begin{equation}\label{swap1}
 \fid = \sum_{\alpha, \beta = 1}^d |\alpha \beta\rangle \langle \beta \alpha|\,, \qquad 
 \gid = \sum_{\alpha, \beta = 1}^d |\alpha \alpha\rangle \langle \beta \beta|\,,
\end{equation}
where $\{ |\alpha\rangle ; \alpha = 1, \cdots d\}$ is the canonical basis for the subsystem $A$ and $B$ (both assumed to be $d$-dimensional).
The first is nothing but the swap operator on the composite system $AB$, while the second is a projector into the (not normalized) maximally entangled state with constant coefficients.
They are related via partial transpose (i.e. $\fid = \gid^T$) and  
are QEWs which we can call of 
{\em Werner class} 
and of {\em Isotropic class} respectively,  
 for reasons which will be clear soon.
Their optimal bounding functions  can be computed  for several
entanglement measures including entanglement of formation ($\eof$), 
relative entropy of entanglement \cite{VOLL}, and convex-roof extended negativity \cite{LEE}.  For the sake of simplicity however in the following we focus only on the former.
In this case  the optimal bounding functions for $\fid$ and $\gid$ are respectively given by
\begin{align} 
 \fwer (x) =& \left\{ 
 \begin{array}{ll}
  h_2 \left(\tfrac{1+\sqrt{1-x^2}}{2} \right) & \mbox{for  $x \leq 0$,} \\ 
  0 &\mbox{for  $x > 0$,} 
 \end{array} \right. \nonumber \label{co}\\ 
 \\ \nonumber 
 \fiso (x) =& \left\{ 
\begin{array}{ll}  
 co[h_2 (\gamma) + (1-\gamma)\log(d-1)]& \mbox{for $x>1$,} \\
 0 & \mbox{for $x\leq 1$,}
 \end{array} \right.
\end{align}
where $ h_2 (y) = - y \log y - (1-y) \log (1-y)$ is the binary Shannon entropy, 
$\gamma$ is the function $\gamma(x) = \frac{1}{d^2} ( \sqrt{x} +
 \sqrt{(d-1)(d-x)} )$, and '$co$' means taking the convex-hull
of the inner expression, i.e.  the largest convex curve nowhere
larger than the given one -- notice however that in (\ref{co}) the argument 
is already convex when $d=2$ \cite{NOTA22}.

Proving that the operator $\fid$ is a QEW with optimal bounding function (\ref{co}) involves the  
knowledge of the properties of Werner states \cite{VOLL,WERNER} and notion of {\em twirling} transformations \cite{BENNETT}.
Werner states $\rho_W$ in an $\mathcal{H}_d \otimes \mathcal{H}_d$ space
are those which are unvariant under the whole $U \otimes U$ group of transformations
\begin{equation}
 (U \otimes U) \rw (U \otimes U)^{\dagger} = \rw,
\end{equation}
for all unitaries $U$ on $\mathcal{H}_d$. They form a one-parameter manifold 
of states which can expressed as 
\begin{equation}
 \rw (f)= \frac{1}{d(d^2-1)}
 \left[
  (d-f)\id + (fd-1)\fid
 \right],
\end{equation}
where $\id$ is the $d^2 \times d^2$ identity operator and where $\fid$ is the swap operator (\ref{swap1}).
The parameter $f$ takes values between -1 and 1 to ensure positivity,
and it coincides with the expectation value of $\fid$ over $\rho_W(f)$, i.e.
\begin{equation} \label{effe}
 f= \trace{\fid \,\rw(f)}.
\end{equation}
The entanglement of formation $\eof \left(\rw(f) \right)$  for the state $\rho_W(f)$ is a known quantity \cite{VOLL}   which can be expressed in terms of the function 
(\ref{co}), i.e. 
\begin{equation} \label{eq:one}
 \eof \left(\rw(f) \right) = \fwer(f) = \fwer\left(\trace{\fid \,\rw}\right).
\end{equation}
To demonstrate that $\fwer$ is a valid bounding function for $\fid$
let us define the following completely-positive trace-preserving mapping,
acting on a generic density matrix $\rho$ of $AB$, 
\begin{equation} \label{eq:twirlwer}
 \tw(\rho) = \int_{\Lambda} d\mu(U)\; (U \otimes U)\, {\rho} \, (U \otimes U)^{\dagger},
\end{equation}
where $\Lambda$ is the unitary group of $\mathcal{H}_d$ and
$d\mu$, its Haar measure, is uniquely defined by compactness of $\Lambda$.
The transformation ${\cal P}_{\Lambda}$ is often labeled as \emph{twirling operation} \cite{BENNETT}.
Three properties of $\tw$ are  important to mention for our purposes \cite{BENNETT,VOLL,LEE}: first,
$\tw$ is a projector, i.e. $\tw^2 = \tw$, and its range is exactly
the 1D manifold of Werner states. Moreover,
$\tw$ preserves the expectation value of $\fid$, i.e.
$ \trace{\fid \,\rho} =  \trace{\fid \,\tw(\rho)}$, which, along with \eqref{effe}, tells us that
\begin{equation} \label{eq:two}
 \tw(\rho) = \rw\left(\trace{\fid \,\rho}\right).
\end{equation}
Finally, since its Kraus decomposition \eqref{eq:twirlwer} is made of tensor product operators, $\tw$ is a
(non-invertible) LOCC transformation. Therefore it can only decrease entanglement, thus providing the following inequality
\begin{equation} \label{eq:three}
 \eof( \tw(\rho)) \leq \eof (\rho).
\end{equation}
Putting together \eqref{eq:one}, \eqref{eq:two}
and \eqref{eq:three} we finally obtain \cite{burkardloss,BENNETT,LEE,VOLL}
\begin{equation} 
 \eof( \rho) \geq \fwer(\trace{\fid \,\rho})\;,
\end{equation}
for any given state $\rho$, which is the proper definition of bounding function for $\fwer$. 
To check optimality,
it is sufficient to see that for any $x \in [-1, 1]$ there is a density matrix
$\rho$ with $\trace{\fid \,\rho} = x$ for which equality holds: this happens for
the Werner state $\rw(x)$. Then, since for any function $\foo$ higher in $x$ than $\fwer$,
the condition \eqref{eq:qew} would be broken, $\fwer$ must be optimal.
\vspace{1em}

The proof of  the optimality of the function (\ref{co})  for the Isotropic QEW $\gid$  is  similar to
the one we exploited in the Werner case. This time we consider isotropic states \cite{HORO,TERHAL,VIDAL,RUNGTA,VOLL},
i.e. the set of states $\rho_I$ invariant under every transformation of the form
\begin{equation}
(U \otimes U^{*}) \rho_I (U \otimes U^{*})^{\dagger} = \rho_I,
\end{equation}
where ``$\vphantom{a}^{*}$" denotes complex conjugation of the matrix entries, say, with respect 
to the canonical basis. Isotropic states form
again a one-parameter family, defined by 
\begin{equation}
 \ri (g)= \frac{1}{d(d^2-1)}
 \left[
  (d-g)\id + (gd-1)\gid
 \right],
\end{equation}
with $0 \leq g \leq d$ and $\gid$ as in (\ref{swap1}). 
The parameter $g$ is also the expectation value
$g = \trace{\gid \,\ri(g)}$. On top of that, their entanglement of formation is
known \cite{TERHAL} to be given by $\eof(\ri(g)) = \fiso(g)$.
Again, we exploit the properties of a twirling operation, this time of the form
\begin{equation}
 \tw'(\rho) = \int_{\Lambda} d\mu(U)\; (U \otimes U^{*})\, \rho \, (U \otimes U^{*})^{\dagger},
\end{equation}
which projects \cite{VOLL,LEE} any given state $\rho$ in the isotropic state $\ri(\trace{\gid \,\rho})$ while
preserving  the expectation value of $\gid$ and degrading the entanglement of formation.
Following these considerations we get \cite{Albe,Brand}
\begin{equation} 
 \eof( \rho) \geq \fiso(\trace{\gid \,\rho}),\label{isoine}
\end{equation}
so that $\fiso$ is a bounding function for $\gid$, and it is optimal
since any function pointwise higher would break the bounding condition \eqref{eq:qew}
upon one isotropic state at least (it is worth pointing out that  the inequality (\ref{isoine}) was originally derived in \cite{Albe}
without invoking the properties of  twirling). 

\section{Local decomposition of QEW} \label{sec:deco}

A practical problem in entanglement detection 
is how to
acquire expectation values of non-local observables (for instance our QEWs),
while we are typically allowed to perform  only LOCC measurements
\cite{PREP,MACCH1,MACCH,SAN,TERHAL2,FILIP,PITTENG,THESIS}. The standard solution is to expand
the witness operator $\qid$ into a sum of local observables of the form 
\begin{equation}
\qid = \label{deco}
\sum_i c_i A_i \otimes B_i \;, 
\end{equation}
 with $c_i$ being some complex coefficients.
The expectation value of $\qid$ can then
be reconstructed by simple data post-processing from the expectation values of 
the tensor product  observables $A_i \otimes B_i$ which can be acquired by performing classical correlated measures on
$A$ and $B$.  
The decomposition (\ref{deco})  is not unique and, unless $\qid$ admits a set of separable eigenvectors \cite{NOTA1},    
it will involve  a certain number of non-commuting operators $A_i\otimes B_i$. 
 In this case several 
independent experimental setups need to be prepared, one for each non-commuting set of 
observables $A_i\otimes B_i$  entering into (\ref{deco}). 
It is thus reasonable to try and
optimize the local decomposition in a way that it corresponds to the smallest possible experimental effort. 
This problem  
gathered a lot of attention \cite{MACCH1, MACCH, FILIP, PITTENG, THESIS} but it appears that  
no general solution has been found yet. 
Instead, several specific solutions for particular classes of witnesses have
been proposed.

In this section we  focus on a local decomposition of the QEWs of the Werner 
and Isotropic class which was first discussed in \cite{MACCH1}. Such solution is  not 
the most efficient one  in terms of number of setup re-preparation.
Indeed it requires $\sim 2 d$ independent setups, while a theoretical lower bound of $d$ was presented in \cite{THESIS} which,
at least for $d$ is prime is known to be achievable \cite{PITTENG}.
Yet the proposed decomposition  presents at least two advantages.
First, as we shall see in Section \ref{multiple}, by simply  reprocessing the data
accumulated  during the various experiment runs, one can  obtain expectation values of an (exponentially large)  class of 
QEW operators which are LOCC-equivalent in the sense of (\ref{eq:questa}) to the one we started with.  
This leads us to the experimental acquisition  of several entanglement bounds which are optimal in the sense of (\ref{eq:optimal}): upon taking their highest value we can hence provide an estimation
of the quantity $q_*$ of (\ref{eq:questa1}). 
Secondly, as we shall see in Section \ref{sec:AS}, the proposed decomposition scheme appears to be naturally suited for 
multi-rail encoding implementations.

\begin{figure} 
 \begin{center}
  \includegraphics[width=\columnwidth]{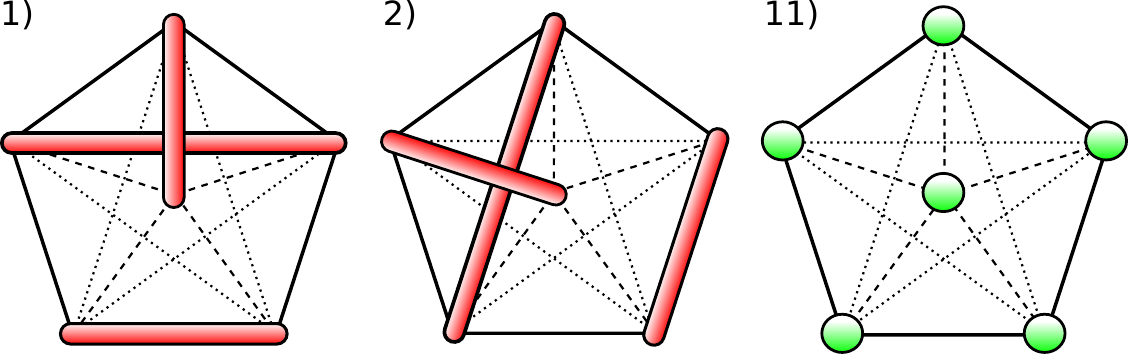}
 \end{center}
 \caption{(colour online) Sketch of the minimization strategy for $d=6$. Turns 1, 2 and 11 are drawn.
 The six-th vertex has been placed in the geometrical center of the pentagon. \label{img:sei}}
\end{figure}

 To start we  introduce the following set of  (local) single-site observables
\begin{equation} \label{eq:pairings}
 \left\{
 \begin{aligned}
  P_{\alpha} &= |\alpha\rangle \langle \alpha|\;, \\
  X_{\alpha \beta} &= |\alpha\rangle \langle \beta| + |\beta \rangle \langle \alpha|\;, \\
  Y_{\alpha \beta} &= i|\alpha\rangle \langle \beta| - i|\beta \rangle \langle \alpha|,
 \end{aligned}
 \right.
\end{equation}
defined for any $\alpha, \beta \in \{0, \cdots, d\}$, $\alpha \neq \beta$ (as before
the vectors $|\alpha\rangle$ stands for the canonical basis of the site).
Since  they form a basis for the space of operators acting on $\mathcal{H}_d$, 
we can express  $\fid$ and $\gid$ as follows, 
\begin{equation} \label{eq:fdec}
 \begin{aligned}
  \fid &= \sum_{\alpha} P_\alpha \otimes P_\alpha + \frac{1}{4} \sum_{\beta \neq \alpha}
   \left( X_{\alpha \beta} \otimes X_{\alpha \beta} + Y_{\alpha \beta} \otimes
   Y_{\alpha \beta} \right), \\
  \gid &= \sum_{\alpha} P_\alpha \otimes P_\alpha + \frac{1}{4} \sum_{\beta \neq  \alpha}
   \left( X_{\alpha \beta} \otimes X_{\alpha \beta} - Y_{\alpha \beta} \otimes
   Y_{\alpha \beta} \right),
 \end{aligned}
\end{equation}
which is of the form (\ref{deco}) with  $d^2$ terms and with the coefficients $c_i$ being real.  
Therefore the expectations values of $\fid$ and $\gid$ can both be acquired with at most order $d^2$ independent
experimental setups. Of course, some of the operators appearing in the sums (\ref{eq:fdec}) commute with each other, so 
in principle it is possible to evaluate them within the same experiment.
For instance any pair of operators of the set
$\Sigma = \{ P_{\alpha}, \,X_{\alpha \beta}, \,Y_{\alpha \beta}  \}$
commute \emph{iff} all the indexes are different, and clearly this commutation
rule extends to all the observable of the form $A \otimes A$ with $A \in \Sigma$.
Exploiting this simple property one can  indeed reduce \cite{MACCH1}
 the number of setup re-preparations needed to obtain the all values 
$\langle P_{\alpha} \otimes P_{\alpha} \rangle$,
$\langle X_{\alpha \beta} \otimes X_{\alpha \beta} \rangle$
and $\langle X_{\alpha \beta} \otimes X_{\alpha \beta} \rangle$ from $d^2$ 
to order $d$
(here we use $\langle \cdots \rangle$  to represent the expectation value with respect to the state $\rho$). 
For the sake of clarity, we review this result by re-deriving using a different method than that presented in \cite{MACCH1}.  To do so we find it convenient to 
 turn the problem into a graph colouring game.
Let us assume that we have a graph made of $d$ vertices and all the possible
edges among them (also known as \emph{complete} graph),
e.g. see figures \ref{img:sette} and \ref{img:sei}; we are also given one green brush to colour the vertices with,
alongside with one blue and one red brushes to paint the edges of the graph.
The idea is that when we paint the $\alpha$-th vertex green we are actually acquiring
$\langle P_{\alpha} \otimes P_{\alpha} \rangle$, while if we paint
in red (resp. blue) the edge connecting the nodes $\alpha$ and $\beta$ we evaluate
$\langle X_{\alpha \beta} \otimes X_{\alpha\beta} \rangle$
(resp. $\langle Y_{\alpha \beta} \otimes Y_{\alpha \beta} \rangle$).
As a rule, during a single turn the player can paint everything
he wants, but \emph{not}:
\begin{itemize}
 \item an edge and a vertex at one of its ends,
 \item two consecutive edges,
 \item the same edge twice;
\end{itemize}
clearly, these game rules encode the commutation relations we discussed.
The aim of this game is to paint all the vertices green and
all the segments in both red and blue spending the minimal number of turns.
Optimal solution for the game are presented next: 
\begin{figure}
 \begin{center}
  \includegraphics[width=\columnwidth]{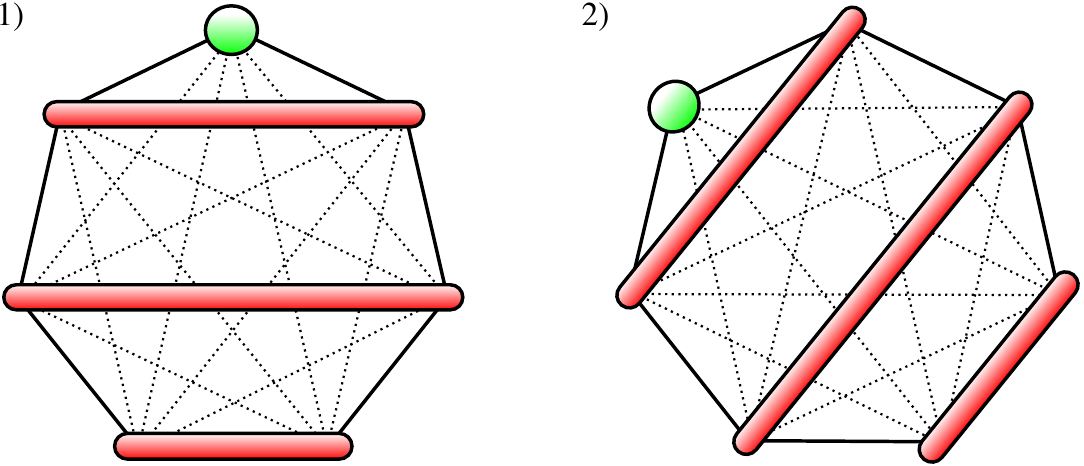}
 \end{center}
 \caption{(colour online) Sketch of the minimization strategy for $d=7$, first two steps.
 A red line connecting vertices $\alpha$ and $\beta$
 means measuring $X_{\alpha \beta} \otimes X_{\alpha \beta}$, the green dot is the measurement
 of $P_{\alpha} \otimes P_{\alpha}$.  \label{img:sette}}
\end{figure}

\emph{Colouring for $d$ odd} - 
For simplicity, we draw the graph like a regular polygon with $d$ edges, including all the diagonals, see figure \ref{img:sette}.
At the first turn the player paints in red one edge and all the diagonals that are parallel to
that edge, as well as the farthest vertex, obviously green.
From turn 2 to $d$ he chooses everytime another edge of the polygon and repeats the process, until
all the edges have been chosen. From turn $d+1$ to $2d$ he does the same thing he did in the
first $d$ turns but using the blue brush instead of the red one.
That completes the game, in $2d$ turns.
This colouring strategy is optimal because, due to the rules, every turn a player cannot paint
more than $(d-1)/2$ segments, and we have $d(d-1)$ segments to paint, so that $2d$
is actually a lower bound for the game turns.  

\emph{Colouring for $d$ even} -
This time we draw the graph as a $d-1$ sided regular polygon, with the additional vertex
in the geometric center, and consider all the edges, diagonals, and vertex-centre segments, see figure \ref{img:sei}.
At the first turn the player paints in red one edge, the parallel disgonals, and the segments
joining the center with the farthest vertex. Afterwards, he rotates the pattern;
and subsequently he does the same withthe blue brush. When he is finished, he uses
a turn to paint all the vertices.
That completes the game in $2d-1$ turns.
As before, $2d-2$ is a lower bound for painting the segments, but if that is the case we need at least one
additional turn to paint the dots. Therefore, this colouring strategy is optimal.
\vspace{1em}

\subsection{Multiple bounds from data processing} \label{multiple}
In the previous section we presented a strategy for
 evaluating
$\langle \fid \rangle$ and $\langle \gid \rangle$ according to the local 
decompositions (\ref{eq:fdec}). Now we want to show that it is possible to reprocess
the same (classical) data to obtain expectation values of other QEWs.
 These are generated from $\fid$ and $\gid$
by means of local invertible transformations as discussed in section \ref{sec:QEW}.
Specifically we focus on the following phase shift operators
\begin{equation}
 W(\xi)= \sum_{j} \xi(\alpha)\; |\alpha\rangle\langle \alpha|\;,
\end{equation}
where $\xi:\{1,\cdots, d\} \rightarrow \{-1,1\}$ is a generic 2-valued function.
Now, let us define
\begin{equation}
\begin{aligned} \label{unitary}
 \fid(\xi) &= \left( \id \otimes W(\xi) \right) \fid \left( \id \otimes W^\dag(\xi) \right) \;,\\
 \gid(\xi) &= \left( \id \otimes W(\xi) \right) \gid \left( \id \otimes W^\dag(\xi) \right)\;.
\end{aligned}
\end{equation}
For all choices of $\xi$, these observables are
QEWs of Werner and Isotropic class respectively (they are just associated with Werner and Isotropic states
defined on a different basis).  Furthermore since they are LOCC equivalent to the original $\fid$ and $\gid$, their
optimal bounding function are still  given by (\ref{co}). By optimizing with respect to all $W(\xi)$
we can write 
\begin{equation} \label{ffd1111}
\begin{aligned}
 \eof( \rho) &\geq& \max_{\xi} \{ \fwer(\trace{\fid(\xi) \,\rho})\}\;, \\
 \eof(\rho)&\geq&   \max_{\xi} \{ \fiso(\trace{\gid(\xi) \,\rho})\}\;, 
\end{aligned}
\end{equation}
where the maximum is taken over all possible choices of $\xi$. These expressions can be further simplified by exploiting
the monotonicity properties of $\fwer(x)$ and $\fiso(x)$  (the former is non increasing while the latter is non decreasing)
to move the optimization over $\xi$ on their arguments.  This yields
\begin{equation} 
 \eof( \rho) \geq
 \fwer(f_*) \;,  \qquad 
  \eof(\rho)\geq   \fiso(g_*) \;, \label{ffd1111333}
\end{equation}
 where we defined the quantities
 \begin{equation} \label{eq:star} \begin{aligned} 
 f_*(\rho) \equiv& \min_{\xi} \{ \trace{\fid(\xi) \,\rho} \}
 \;,  \\
 g_*(\rho) \equiv&   \max_{\xi} \{\trace{\gid(\xi) \,\rho}\} \;.
\end{aligned} \end{equation}
These are complicated (non linear) function of the state $\rho$ which are related to the quantity $q_*$ 
of (\ref{eq:questa1}). The main difference is that the latter was optimized with respect to the whole
set of unitary LOCC transformations $\Gamma$, while $f_*$ and $g_*$ are optimized only with respect to the a
proper subset of such set, namely
the phase flips (\ref{unitary}). 
Luckily enough we can simplify $f_*$ and $g_*$ by rewriting the operators $\fid(\xi)$ and $\gid(\xi)$
in terms of the operators (\ref{eq:pairings}). Indeed by doing so we find that they are
made of the very same $d^2$ operators appearing in \eqref{eq:fdec}, namely, 
\begin{multline} \label{defgamma}
 \fid (\xi) = \sum_{\alpha} P_{\alpha} \otimes P_{\alpha}
 + \frac{1}{4} \sum_{\alpha \neq \beta} \xi(\alpha) \xi(\beta) \; \times \\ \times
 \left( X_{\alpha \beta} \otimes X_{\alpha \beta} + Y_{\alpha \beta} \otimes Y_{\alpha \beta} \right)\;,
\end{multline}
and
\begin{multline} \label{deggamma}
 \gid (\xi) = \sum_{\alpha} P_{\alpha} \otimes P_{\alpha}
  + \frac{1}{4} \sum_{\alpha \neq \beta} \xi(\alpha) \xi(\beta)  \; \times \\ \times
 \left( X_{\alpha \beta} \otimes X_{\alpha \beta} - Y_{\alpha \beta} \otimes Y_{\alpha \beta} \right)\;.
\end{multline}
This implies in particular that their expectation values can be recovered by the same set of LOCC measurements
required for $\langle \fid\rangle$ and $\langle\gid\rangle$, allowing us to   build $2^{d-1}$ independent new Werner QEWs of the form $\fid(\xi)$, as
well as $2^{d-1}$ independent new Isotropic QEWs of the form $\gid(\xi)$ by
simple data post-processing
 [the number $2^{d-1}$ follows from 
the fact that there are $2^d$ possible $\xi$ functions, but
$\fid(\xi) = \fid(-\xi)$ and $\gid(\xi) = \gid(-\xi)$].
Clearly this  is a very simple and cheap way to substantially increase the information 
we can gather on $\rho$.  
As a matter of
fact the expressions (\ref{defgamma})
and (\ref{deggamma}) are also useful to simplify the optimizations of \eqref{eq:star}.
 In particular define  the $d^2\times d^2$ matrix $M^{(\pm)}$ of elements,
\begin{equation}
 \begin{aligned}
  M_{\alpha,\alpha}^{(\pm)} =&   \langle P_{\alpha} \otimes P_{\alpha}\rangle\;, \\
  M_{\alpha,\beta}^{(\pm)} = &\frac{1}{4} ( \langle X_{\alpha \beta} \otimes X_{\alpha \beta}\rangle \pm \langle
  Y_{\alpha \beta} \otimes Y_{\alpha \beta} \rangle )\;, 
  \; \alpha \neq \beta
 \end{aligned}
\end{equation}
where, as usual, $\langle \cdots \rangle$ stands for taking the expectation value with respect to $\rho$. These matrices
are composed by the measurement outcomes one acquires when recovering 
the expectation values of $\fid$ and $\gid$ via the local decomposition introduced in the previous section. In particular we can write,
\begin{equation} \label{fvv}
 \begin{aligned}
\trace{\fid(\xi) \,\rho} &=  \sum_{\alpha, \beta}
  \xi(\alpha)  \; M_{\alpha,\beta}^{(+)}  \;  \xi(\beta) \equiv \vec{\xi}^{\;T} \cdot M^{(+)} \cdot \vec{\xi} \;, \\
 \trace{\gid(\xi) \,\rho} &= \sum_{\alpha, \beta}
  \xi(\alpha)  \; M_{\alpha,\beta}^{(-)}  \;  \xi(\beta) \equiv {\vec{\xi}}^{\;T} \cdot M^{(-)} \cdot \vec{\xi} \;,
 \end{aligned}
\end{equation}
where $\vec{\xi} \equiv ( \xi(1), \cdots, \xi(d))^T$ is the vector formed by the output entries of the binary function $\xi$. 
Therefore, in order to determine the values (\ref{eq:star}) is sufficient
to respectively minimize or maximize these quantities over $\vec{\xi}$; i.e.
\begin{equation} \label{ffd5} \begin{aligned} 
 f_*(\rho) &\equiv \min_{\vec{\xi}} \; \vec{\xi}^{\;T} \cdot M^{(+)} \cdot \vec{\xi}
 \;,  \\
 g_*(\rho) &\equiv   \max_{\vec{\xi}}\;  \vec{\xi}^{\;T} \cdot M^{(-)} \cdot \vec{\xi} \;.
\end{aligned} \end{equation}
 Unfortunately, apart from the trivial case of $d=2$, providing a general analytical solution for these expressions is
 nontrivial and dependent on $M^{(\pm)}$ (indeed it is formally equivalent to find a ground state of
 a frustrated, classical many-body spin system). 
On the other hand, one can easily evaluate upper bounds as follows
\begin{equation} \begin{aligned}
  f_* (\rho) &\geq \sum_{\alpha} M^{(+)}_{\alpha, \alpha}
  - \sum_{\alpha \neq \beta}
  | M^{(+)}_{\alpha, \beta} | \;,
  \\ \label{5}
  g_*(\rho) &\leq  \sum_{\alpha} M^{(-)}_{\alpha, \alpha}
  +  \sum_{\alpha \neq \beta}
  | M^{(-)}_{\alpha, \beta} | \;,
\end{aligned} \end{equation}
even  though for $d>3$  they will be in general not tight and will not be useful to
lower bound the entanglement of formation of the system.
Luckily enough however the optimizations (\ref{ffd5}) are  treatable numerically, especially considering that for the experimental implementation we 
are interested in,  $d$ is usually of the order of magnitude of units (see next section).

\section{Implementation: multi-rail encodings in quantum optics}\label{sec:AS}

\begin{figure}
 \begin{center}
  \includegraphics[width=200pt]{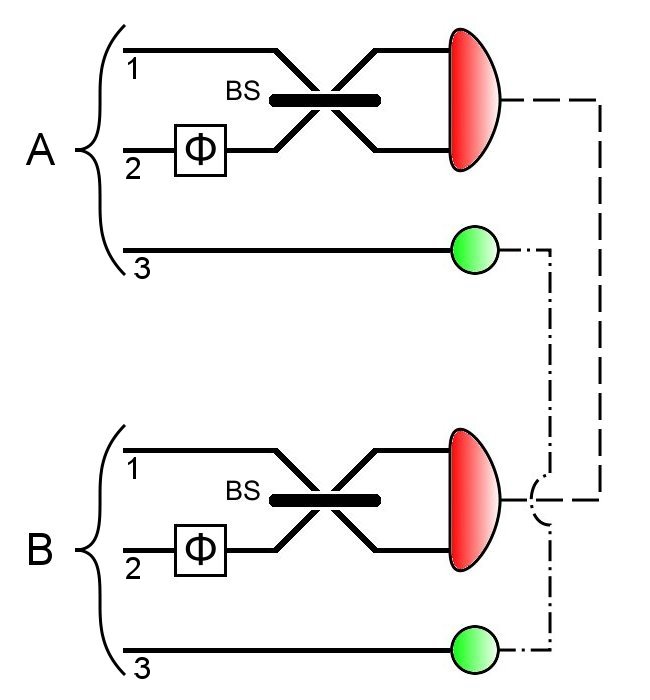}
 \end{center}
 \caption{ \label{img:setup}
 (colour online) Example of data sampling for $AB$ system in which the local degree of freedom of each site is composed by three independent optical channels (here represented as spatially 
 separated for simplicity).
 The figure represent the fist step of the detection strategy: 
 Applying the $50\%$ transmittive beam splitter and measuring
 $\langle(N_{A,1} - N_{A,2})(N_{B,1} - N_{B,2}) \rangle$
 gives us the expectation value of 
 $X_{1,2} \otimes X_{1,2}$ (if we put
 a $\Phi = \pi / 2$ phase shift on channel 2, we would have
 $Y_{1,2} \otimes Y_{1,2}$ instead); at the same
 time we also acquire $\langle P_{3} \otimes P_{3} \rangle =
 \langle N_{A,3}\, N_{B,3} \rangle $.
 For later steps, we need to permute the channels in the picture
 and perform those measurements as well.
 }
\end{figure}

QEW techniques for estimating
lower bounds of entanglement of a quantum state are suitable, in principle,
to any experimental setting or physical apparatus.
Yet, we are inclined to think that the proposed local decomposition scheme of Werner
and Isotropic witnesses we presented in the previous section would be particularly fit
to \emph{multi-rail encoding} experimental settings.
In such implementations, a state $|\alpha\rangle$ of the canonical basis
is represented by the presence of a particle/carrier in a specific
transmission channel (e.g. spatial mode, polarization, time-bin or orbital angular momentum) indexed by $\alpha$. Multi-rail encodings are common both in optics \cite{multi,HYP1,HYP2,HYP,GAO} and in electronics  (e.g. see \cite{beenakker,BEEN11}).
In the following however we will   focus only on the former case as it is intrinsically more flexible. 
For these architectures the QEW detection strategy discussed in section \ref{sec:deco} entails to properly mixing a pair of 
incoming modes via simple beam-splitter transformations and to perform photo-coincidence counting measurements
at the exit port of the device, see figure \ref{img:setup}.
Accordingly the proposed LOCC measurements appear to be more suited to detect entanglement in these systems than
(say) the optimal schemes of \cite{PITTENG} which requires 
coherent mixing among   multiple channels.

Consider for instance two separated sets $A$ and $B$ of (monochromatic) spatial optical modes. 
Within each of the two subsystems a single photon is injected and allowed to flow through $d$ possible optical paths. In one of experimental setup of \cite{HYP}
for instance, $d=8$ was realized by exploiting polarization and 2 different spatial degree of freedom.
A generic (pure) state of the system can then be  written as 
\begin{equation}
 |\Psi_{in}\rangle =
  \sum_{\alpha, \beta = 1}^{d} \Phi_{\alpha \beta}\;\, \hat{a}^{\dagger}_{A, \alpha}\; \hat{a}^{\dagger}_{B, \beta}
 |0\rangle =\sum_{\alpha, \beta=1}^d \Phi_{\alpha \beta} |\alpha \beta\rangle\;,
\end{equation}
where the $\hat{a}_{j, \alpha}$ is the annihilation operator of a photon
flowing through channel $\alpha$ at port/subsystem $j\in \{ A, B\}$, and where  $|\alpha \beta\rangle$ stands for the two photon state  $\hat{a}^{\dagger}_{A, \alpha}\; \hat{a}^{\dagger}_{B, \beta}
 |0\rangle$. 
We will describe  local transformations on photonic state as a forward elastic scattering process
\begin{equation}
 \hat{b}_{A,\alpha} = \sum_{\gamma = 1}^{d} U_{\alpha, \gamma} \;\hat{a}_{A, \gamma}\;, \qquad
 \hat{b}_{B,\beta} = \sum_{\gamma = 1}^{d} V_{\beta, \gamma} \;\hat{a}_{B, \gamma}\;,
\end{equation}
with $U$ and $V$ unitaries which acts on the $A$ and $B$ sets respectively. 
After applying such transformation all we have to do is record
the channel-selective photo-coincindence measurements 
\begin{equation} \begin{aligned} \label{ph-count}
 \langle \hat{N}_{A,\alpha} \;\hat{N}_{B,\beta} \rangle &=
 \langle \Psi_{in}| \hat{b}_{A,\alpha}^{\dagger} \;\hat{b}_{A,\alpha}\;
 \hat{b}_{B,\beta}^{\dagger} \; \hat{b}_{B,\beta} |\Psi_{in}\rangle
 \\ &= \left| \langle \alpha \beta | U \otimes V |\Psi_{in} \rangle \right|^2.
\end{aligned} \end{equation}
This way we achieve simultaneously the expectation values of the $d^2$ one dimensional
projectors of the form
 $U^{\dagger} \otimes V^{\dagger} \;| \alpha \beta \rangle \langle \alpha \beta |\; U \otimes  V$.
The last step is to identify a proper set of unitaries $U$ and $V$ that lead to the determination of  
the quantities $\langle P_{\alpha} \otimes P_{\alpha}\rangle$, $\langle  X_{\alpha \beta} \otimes X_{\alpha \beta}\rangle$, and $\langle Y_{\alpha \beta} \otimes Y_{\alpha \beta}\rangle$
which are needed for characterizing the selected QEW.
For instance assume we want to measure 
the expectation value of $X_{\alpha \beta} \otimes X_{\alpha \beta}$, where $\alpha$ and $\beta$ represent to two distinct channels. Then $U$ and $V$ could be implemented as Hadamard gates, acting in both subsystems A and B, which couples the pair of such channels.
This entails the implementation of a beam-splitter transformation of transmissivity $50\%$ which coherently mixes the two channels.
After that, we can convert coincidence measurements into data via
\begin{equation}
 \langle X_{\alpha \beta} \otimes X_{\alpha \beta} \rangle =
 \langle ( \hat{N}_{A, \alpha} - \hat{N}_{A, \beta} )
 ( \hat{N}_{B, \alpha} - \hat{N}_{B, \beta} ) \rangle.
\end{equation}
Similarly, if our goal were to obtain
$\langle Y_{\alpha \beta} \otimes Y_{\alpha \beta} \rangle$ instead, 
the unitarities $U$ and $V$ could be chosen as
Hadamard transformations, this time
applied after a local phase shift (e.g. of $\pi/2$ acting upon channel $\beta$).
See figure \ref{img:setup} for details.
We can now translate into experimental operations every 
possible strategy for the local QEW decomposition of section \ref{sec:deco}.

As a concluding remark we point out that the same analysis 
can be generalized to an electronic implementation of multi-rail encodings \cite{BEEN11,beenakker}. 
In this case the ports $A$ and $B$ will correspond to independent leads, whose transverse modes play the role of the optical paths.
Under appropriate conditions \cite{beenakker} the coincidence counting
(\ref{ph-count}) can be related to the zero-frequency shot noise, which is a measurable quantity.
In \cite{faoro}, for example, a procedure for constructing EWs was employed for different solid-state systems for the case of two transverse modes per lead.
In such systems, however, an experimental limitation is represented by the difficulty of working with more than two modes.

\section{Conclusions}

We have reviewed the notion of QEWs and clarified that there exist a class of such operators for which
the optimal bounding function can be easily determined without necessarily 
going through the complex optimization procedure of (\ref{eq:optimal}) and (\ref{alt}). 
For these QEWs we then considered an explicit local decomposition that allows one to acquire the needed expectation values via a limited number of 
LOCC measurements. The proposed scheme is not optimal in terms of number of independent setups one need to prepare, but
has two main advantage with respect to other schemes \cite{PITTENG}. In particular we can easily optimize with respect a large class
of unitary LOCC providing an operational characterization of the associated non linear quantities, see (\ref{fvv}).
Furthermore we have shown that it is suitable for experimental implementations based on multi-rail encodings by means of simple two-mode transformations (beam-splitter mixings) followed by 
correlated photo-counting detections.

\vspace{1em}
\emph{Acknowledgements - }
We would like to thank P. Hyllus, C. Macchiavello, P. Mataloni, and A. Osterloh, for useful comments and discussions. We are particularly grateful to P. Hyllus for pointing out
\cite{THESIS}.
This work was in part supported by the FIRB IDEAS project RBID08B3FM.


\begin{thebibliography}{1}
\bibitem{NOTA11} 
The QEW set is clearly invariant under the whole set of LOCC transformations. However if the mapping $\Gamma$ is
not invertible (i.e. not unitary) then the optimal bounding function of the new observable $\Gamma^{\ast} (\gid)$ 
needs not to coincide with the original one.
\bibitem{NOTA22} 
This is a rather special case where one can easily show that
the two QEW are basically LOCC equivalent in the sense discussed in (\ref{eq:questa}).
Indeed in this case  $\gid$ can be expressed in terms
of $\fid$ via a proper local unitary transformation plus a constant term which can be compensated via
classical data post-processing.
\bibitem{NOTA1} 
This case however is not relevant as an operator of that form wouldn't be able to detect entanglement.
\end{thebibliography}
\end{document}